\documentclass[12pt]{article}
\usepackage{amssymb,graphicx}
\usepackage{epstopdf}
\usepackage{amsmath,amsfonts}
\usepackage{epsfig}
\usepackage[titletoc,title]{appendix}
\usepackage{xcolor}
\numberwithin{equation}{section}

\def\d{\partial}

\newcommand{\be}{\begin{equation}}
\newcommand{\ee}{\end{equation}}
\newcommand{\bea}{\begin{eqnarray}}
\newcommand{\eea}{\end{eqnarray}}
\newcommand{\bg}{\begin{gather}}
\newcommand{\eg}{\end{gather}}
\newcommand{\bseq}{\begin{subequations}}
\newcommand{\eseq}{\end{subequations}}

\renewcommand{\ln}{\mathop{\rm ln}\nolimits}

\newcommand{\qq}[1]{``#1''}
\newcommand{\mb}{_{\mu}}
\newcommand{\mo}{^{\mu}}

\newcommand{\mnb}{_{\mu\nu}}

\newcommand{\gl}{\big(}
\newcommand{\gr}{\big)}
\newcommand{\bel}[1]{\be\label{#1}}
\newcommand{\deta}{\d_\eta}

\begin{document}
\begin{flushright}
\end{flushright}
\vspace{10pt}
\begin{center}
  {\LARGE \bf Geodesic (in)completeness \\[0.3cm] in general metric frames} \\
\vspace{20pt}
V. A. Rubakov$^{a}$, C. Wetterich$^{b}$\\
\vspace{5pt}
$^a$\textit{
Institute for Nuclear Research of
         the Russian Academy of Sciences,\\  60th October Anniversary
  Prospect, 7a, 117312 Moscow, Russia}\\
\vspace{5pt}
$^b$\textit{
Institute for Theoretical Physics, Heidelberg University, Philosophenweg 16, 69120 Heidelberg, Germany}
\end{center}
\vspace{5pt}

\begin{abstract}

The geometric concept of geodesic completeness depends on the choice of the metric field or \qq{metric frame}. We develop a frame-invariant concept of \qq{generalised geodesic completeness} or \qq{time completeness}. It is based on the notion of physical time defined by counting oscillations for some physically allowed process. Oscillating solutions of wave functions for particles with varying mass permit the derivation of generalised geodesics and the associated notion of completeness. Time completeness involves aspects of particle physics and is no longer a purely geometric concept.

\end{abstract}
\section{Introduction}

The lack of geodesic completeness~\cite{RP},~\cite{SH} is often considered as an indication that a given cosmological model needs some extension in order to cure this shortcoming. Many models of inflationary cosmology are geodesically incomplete. The corresponding general theorems~\cite{Borde:2001nh, MV} are typically based on Einstein gravity (general relativity) coupled to canonical scalar fields and other particles with field-independent fixed mass. Many recent models for very early cosmology involve, however, scalar fields whose coupling to particles induces field-dependent masses. They also often have a non-minimal coupling to gravity or kinetic terms involving more than two derivatives. This challenges the meaning of geodesic completeness for such models.

If masses of particles depend on scalar fields, similar to the Higgs mechanism in the standard model, their motion is no longer along geodesics. The field-dependence of the mass induces scalar-field mediated forces in addition to the gravitational force. For a non-minimal gravitational coupling of scalar fields, as realized for dynamical dark energy or Higgs inflation, a Weyl transformation to the Einstein frame changes the geometry and can lead to field-dependent particle masses. This raises the general question in which metric frame geodesic completeness should be evaluated. In particular, \qq{genesis models} with flat Minkowski geometry in the infinite past~\cite{CNT, CHK, MRV1, MRV2, Creminelli:2016zwa, AMTZ} are obviously geodesically complete. After a Weyl transformation to the Einstein frame geodesic completeness may be lost. On the other hand, typical geodesically incomplete inflationary models can be transformed by a Weyl transformation to a geodesically complete \qq{primordial flat frame}~\cite{Wetterich:2020fth} with flat Minkowski geometry in the infinite past. In quantum field theory physical observables do not depend on the choice of fields used to describe them. They are independent of the choice of the metric frame.

We infer that a purely geometric notion of geodesic (in)completeness is not sufficient for a judgment if a given cosmological model is acceptable or not. We propose to replace the geometric notion of geodesic completeness by a concept of \qq{generalised geodesic completeness} or \qq{time completeness} that does not depend on the choice of the metric frame. It is based on the notion of \qq{physical time} defined by counting oscillations for some physically possible process.

It is straightforward to demonstrate the issue explicitly. In non-Euclidean geometry with Lorentz signature --- the type of geometry
relevant for describing physical space-time --- the standard criterion of past geodesic
completeness of (a patch of) a manifold with spatially flat homogeneous and isotropic metric
\be
ds^2 = dt^2 - a^2(t) d{\bf x}^2 =
a^2 (\eta) (d\eta^2 - d{\bf x}^2)
\label{aug28-22-60}
\ee
is~\cite{Borde:2001nh}
\be
\int_{-\infty}^t~a(t)~dt = \infty\ .
\label{aug17-22-1}
\ee
In the opposite case of a convergent integral in \eqref{aug17-22-1}, or if the initial time cannot be taken to $-\infty$, the
 (patch of a) manifold is said to be geodesically incomplete.

An important notion for understanding the physics beyond the purely geometrical aspects is the metric
frame (conformal frame). A Weyl (conformal)
scaling relates two metric tensors,
\be
  g_{\mu \nu} = \Omega^2 (x) \hat{g}_{\mu \nu}\ ,
\label{nov21-18-20}
  \ee
  where $\Omega (x)$ is typically (but not necessarily)
  a function of some field $\chi$
  (say, dilaton). This Weyl scaling transforms one
  metric frame to another. A standard example is the Jordan frame vs Einstein frame in scalar-tensor
  theories. Since eq.~\eqref{nov21-18-20} can be viewed merely as a
  field redefinition, physics in different metric frames should be
  the same\footnote{Modulo subtle cases explored, e.g., in
    Ref.~\cite{Bars:2011aa}.} if all quantities as particle masses or temperature are transformed properly~\cite{CWQ1}. This holds even though the geometric picture may look
  quite different, as emphasized again recently, e.g. in
  Refs.~\cite{Bars:2011aa,Behnke:2000um,Wetterich:2013aca}.

  The integral in \eqref{aug17-22-1} is not invariant under the Weyl scaling
  \eqref{nov21-18-20}: in terms of conformal time, this integral
  reads $\int\, a^2(\eta)\, d\eta$, while under the Weyl scaling one has
  $a(\eta) = \Omega (\eta) \hat{a}(\eta)$. By a suitable choice of $\Omega(\eta)$ one can switch between geodesic completeness and incompleteness for the purely geometrical definition~\eqref{aug17-22-1}. The need for a generalisation of the concept of completeness becomes apparent.
  
  \section{Geodesics and physical time for
    varying particle mass.}

  \subsection{Clocks}

  Consider a free scalar field theory in non-Euclidean background
  with action
\be
\mathcal{S} = \int~d^dx Z (x) \sqrt{-g} \left(
\frac{1}{2} g^{\mu \nu} \d_\mu \phi \d_\nu \phi - \frac{m^2(x)}{2} \phi^2
\right) \; ,
\label{aug28-22-2}
\ee
where $Z(x)$ and $m(x)$ may be functions of background fields other than the metric. Under the Weyl scaling \eqref{nov21-18-20} (with $\Omega$ independent
of $\phi$) the action has the
same functional form with
\be
m(x) = \Omega^{-1}(x) \hat{m} (x) \; , \;\;\;\;\; Z(x) = \Omega^{-2} (x)
\hat{Z}(x) \; .
\label{aug28-22-1}
\ee
The definition of physical time, as well as the criterion of generalised geodesic
(in)completeness must be invariant under the combined
Weyl scaling \eqref{nov21-18-20}, \eqref{aug28-22-1}.

With the action given by \eqref{aug28-22-2}, the field equation reads
\be
Z^{-1} \nabla^\mu(Z \nabla_\mu \phi) + m^2(x) \phi = 0 \; .
\label{may12-20-1}
\ee
We are interested in fast oscillating solutions
\be
\phi (x) = A(x) \mbox{e}^{iS (x)}\; ,
\label{sep18-22-1}
\ee
where $S(x)$ obeys the eikonal (Hamilton--Jacobi) equation
\be
g^{\mu \nu} \d_\mu S \d_\nu S - m^2(x) = 0 \; ,
\label{nov21-18-3}
\ee
and $A(x)$ is a slowly varying function. Equation \eqref{nov21-18-3}
is invariant under the Weyl scaling, since both $g^{\mu \nu}$ and
$m^2$ get multiplied by $\Omega^{-2}$. Thus, $S$ is a metric frame
independent quantity.
(Eq.~\eqref{may12-20-1} is
also invariant, so $\phi$ is also frame independent.)

Let $S(x)$ be a certain solution to eq.~\eqref{nov21-18-3}. We ask how
physical time may be defined by counting the number of oscillations of the
wave function $\phi(x)$. For this purpose one needs to define a
trajectory $x^\mu (\sigma ; x(\sigma_0))$ on which
one measures the number of oscillations as one moves along the trajectory
as $\sigma$ increases. Here $x^\mu (\sigma_0)$ denotes the ``starting point''
of the trajectory at $\sigma = \sigma_0$. By evaluating $S(\sigma)$
on this trajectory, one counts the oscillations: an oscillation occurs
whenever $S$ increases by $2\pi$. Thus, $S(\sigma)$ defines a ``clock''
whose ``ticks'' count the oscillations along a given trajectory.
A family of trajectories $x^\mu (\sigma   ; x(\sigma_0))$ parametrized by
$x^\mu (\sigma_0)$ defines a reference frame in which the physical
time is measured. Thus, our concept of physical time
  introduces in a natural way the relativistic notion of a reference frame
  in quantum field theory and more general probabilistic
  systems~\cite{CWPW}. Physical time depends on the
particular solution $S(x)$, as well as on the choice of reference frame:
for given $S(x)$ there is a large family of clocks corresponding to
different trajectories/reference frames.


At this point the definition of physical time seems to be highly arbitrary, depending on the choice of clocks. Many clocks are equivalent, however, since
physical time measured by different clocks
can be gauged. Two clocks defined by different wave functions $S(x)$ and/or different reference frames are considered as equivalent if physical time in one clock is mapped monotonously and uniquely to the physical time in the
other clock, and a finite (infinite) number of ticks of the
second clock corresponds to a finite (infinite) time interval of the
first clock and vice versa. A family of equivalent clocks constitutes a \qq{clock system}. Physical time can then be considered as a property of a clock system. One can take
a particular clock in the equivalence class to define a ``standard time''.

One possible simple choice of reference frame takes trajectories
for which ${\bf x}$ remains constant and $x^0$ is set equal to $\sigma$.
This may be considered as natural for the Universe which is
homogeneous and isotropic in some average sense, with $x^\mu$
associated with comoving coordinates. We will come back to
  such a ``cosmic reference frame'' in sect.~\ref{sec:CMP}.
and focus first on the concept of physical time for special trajectories,
namely, generalised geodesics. This has the advantage that generalised
geodesics can be constructed for massive particles
with arbitrary metric $g_{\mu \nu}(x)$ and mass $m(x)$. Furthermore,
generalised geodesics can be used
for a general construction of solutions to eq.\eqref{nov21-18-3}.

\subsection{Geodesic equation}

For given $S(x)$, let us define particular ``geodesic trajectories''
$x^\mu (\sigma)$ by
\be
\frac{d x^\mu}{d \sigma} = g^{\mu \nu} \frac{\d S}{\d x^\nu} \; .
\label{sep17-22-1}
\ee
We will establish below that for $S$ obeying 
\eqref{nov21-18-3} these trajectories are generalised geodesics or world lines.
It follows from
\eqref{nov21-18-3} and \eqref{sep17-22-1}
that the parameter $\sigma$ is related to
the interval $\tau$
(which would be proper time if $m(x)$ were constant)
associated with metric $g_{\mu \nu}$ as follows:
\be
d\sigma = \frac{1}{m (x)} \sqrt{g_{\mu \nu}(x) dx^\mu dx^\nu} \equiv \frac{1}{m(x)}
d \tau \; .
\ee
Neither $\sigma$ nor $\tau$ are invariant under the
Weyl scaling \eqref{nov21-18-20}, \eqref{aug28-22-1}.
Indeed, for $d\tau^2 = g_{\mu \nu} (x) dx^\mu dx^\nu$,
proper time transforms as
\be
d\tau = \Omega (x) d \hat{\tau} \; .
\ee
The action (eikonal) along the generalised geodesic trajectory defined by
 \eqref{sep17-22-1} 
is
\be
S(\sigma) - S(\sigma_0)
= \int~dx^\mu~\frac{\d S}{\d x^\mu} = \int~m^2(x)~d\sigma \; .
\label{sep17-22-2}
\ee
Being directly related to a discrete counting of oscillations a time derived from this quantity is the same in all metric frames.

What remains is to find trajectories obeying 
\eqref{sep17-22-1} with $S$ a solution to \eqref{nov21-18-3}. These
are the generalised geodesics which obey
\be
\frac{d u^\mu}{d \sigma} + \Gamma^\mu_{\nu \lambda} (x) u^\nu u^\lambda
- \frac{1}{2} g^{\mu \nu} (x) \frac{\d m^2 (x)}{\d x^\nu} = G^\mu= 0 \; ,
\label{nov21-18-12}
\ee
where we define
\be
u^\mu = \frac{dx^\mu}{d\sigma} = g^{\mu \nu} 
  \frac{\d S}{\d x^\nu} \; .
  \label{oct02-1}
  \ee
It is easy to see that the solution of \eqref{nov21-18-12}
    obeys eq.~\eqref{nov21-18-3}, which can be written with~\eqref{oct02-1} as
    \be
    g_{\mu \nu} u^\mu u^\nu = m^2 (x) \; .
    \label{XA}
    \ee
More precisely, if initial values of $u^\mu$ at $\sigma_0$ are chosen such that $A=u^\mu u_\mu-m^2=0$ ($u_\mu=g_{\mu\nu}u^\nu$), then $A=0$ holds for all $\sigma$ and therefore for the whole trajectory. Taking a $\sigma$-derivative of \eqref{XA} one finds
    \be
    \frac{d m^2}{d \sigma} = \frac{\d m^2}{d x^\nu} u^\nu
    = 2 g_{\mu \nu} u^\nu \frac{d u^\mu}{d \sigma}
    + \frac{\d g_{\mu \nu}}{\d x^\rho} u^\rho u^\mu u^\nu
    \label{XB}
    \ee
    or
    \be
    \left(  \frac{\d m^2}{d x^\nu} g^{\mu \nu} - 2 \frac{d u^\mu}{d \sigma}
    -  \frac{\d g_{\lambda \sigma}}{\d x^\rho} g^{\lambda \mu} u^\rho u^\sigma
    \right) g_{\mu \tau} u^\tau = -2G^\mu u_\mu = 0 \; .
    \label{XC}
    \ee
Thus the generalised geodesic equation~\eqref{nov21-18-12} implies indeed $\d A/\d\sigma=0$.

More generally, all solutions of eqs.~\eqref{nov21-18-3},~\eqref{sep17-22-1} have to obey the geodesic equation~\eqref{nov21-18-12}.
  In Appendix A we derive eq.~\eqref{nov21-18-12} using the
  general approach to first order partial differential equations
  based on the method of characteristics; 
see
Ref.~\cite{Ayaita:2011ay} for an alternative derivation.
Even though eq.~\eqref{nov21-18-12} does not appear invariant under
the Weyl scaling \eqref{nov21-18-20}, \eqref{aug28-22-1}, it actually
can be cast in Weyl invariant form.
This is demonstrated in Appendix~\ref{appB}. Thus, the generalised geodesics or world lines
are the same in all metric frames, as they should.

The generalised geodesics describe the trajectories (world lines)
of point-like particles with varying mass. For $\d m^2 /\d x^\mu \neq 0$ they
are no longer purely geometrical objects. We can attribute
the difference to the geometrical geodesics to the action of a
``fifth force'' induced by fields which are responsible for the $x$-dependence
of $m^2(x)$ \cite{Ayaita:2011ay}.

\subsection{Geodesic physical time}

 With oscillating field $\phi(x)$ given by
 \eqref{sep18-22-1}, one particular physical time is such
 that its
  interval $dT$
is just the
increment of the action along the generalised geodesics
\be
d T \equiv d S = m^2(x) d\sigma = m(x) d\tau \; .
\label{aug28-22-11}
\ee
This geodesic physical time $T$ counts the number
  of oscillations of the wave function in the rest frame of a massive $\phi$-particle
  which moves along the generalised geodesic derived from $S$.
This time $T$
{\it is} invariant under the
Weyl scaling \eqref{nov21-18-20}, \eqref{aug28-22-1}. One interpretation
of $d T$ is that it is proper time interval
measured in units of inverse
particle mass; with this motivation, a similar parameter 
was suggested in Refs.~\cite{Wetterich:2014zta,Bars:2013vba}.

It is possible to define $dT$ uniquely in terms of solutions of the field equation~\eqref{may12-20-1} once
     the dynamics is in the WKB regime.
  Such solutions can be realized in all epochs of the universe, including inflation  or other \qq{beginning epochs}, provided that massive fields exist. (Scalars may be replaced by fermions.) If at a cosmological epoch
  an observer moving on a generalised geodesic builds a clock apparatus using particles with masses proportional to $m(x)$, this apparatus will measure $dT$. (More precisely, the time measured by the apparatus is equivalent to the time defined by the oscillations of the wave function.)

We can now proceed to a generalised definition of geodesic completeness which is independent of the choice of the metric field and applies to fields with varying mass. A setup is past geodesically complete if for fast oscillating solutions of the field equation for a free massive field every world line or generalised geodesic trajectory leads to infinite geodesic physical time
%
from the beginning to time $t$, i.e., for any generalised geodesic
\be
\int_{t=-\infty}^t ~d T = \infty\ . 
\label{aug28-22-5}
\ee
(Free massive particles have ``infinite lifetimes''.) It is
past geodesically incomplete otherwise. It is worth emphasizing
that this notion depends, generally speaking, on the sort of particle:
in a given metric frame, different particles may have different
masses $m(x)$ (say, due to different interactions with background fields
other than metric), so that one sort of particles may have a divergent integral
in \eqref{aug28-22-5}, whereas a similar integral for another sort may be convergent. In other words, geodesic physical times defined by different sorts of particles may be inequivalent
in the sense that the time elapsed from the beginning may be infinite for
one sort and finite for another. 

The generalised geodesic completeness or \qq{time completeness} is no longer a purely geometric  notion. It involves \qq{particle physics} through the function $m(x)$. Abandoning a purely geometric notion is mandatory for a frame invariant formulation. For a given metric frame we can still define the standard geometric notion of geodesic completeness. Except for frames leading to constant $m$ it should be distinguished, however, from the more physical concept of time completeness.

\subsection{Wave function and trajectories}

A given eikonal
  function $S$ generates a family of world lines or generalised geodesics according to~\eqref{sep17-22-1}. Different members of this family correspond to different initial conditions at $\sigma_0$. All members of this family obey the geodesic equation~\eqref{nov21-18-12}. On the other hand, there are many more world lines obeying~\eqref{nov21-18-12} which do not obey the condition~\eqref{sep17-22-1} for the given $S$. For example, we may consider a different solution $S'$ which again generates a family of world lines. The world lines generated by $S'$ do not obey eq.~\eqref{sep17-22-1} for $S$. Nevertheless, we can use world lines generated by $S'$ or, more generally, arbitrary world lines obeying~\eqref{nov21-18-12}, as trajectories on which the physical time defined by $S$ is explored. All these different world lines constitute possible trajectories on which the oscillations of $\phi~\sim e^{iS}$ are counted. They correspond to clocks in different reference frames.

The question arises how the clocks on different world lines are related.
  We demonstrate this issue by a simple example, taking for $\phi$
  a plane wave solution in flat space for constant $m$. We specify
  a given wave function by the
  momentum $p\mb$,
  \be
  \label{G1}
S(x)=p\mb x\mo\ ,\quad p\mo p\mb=m^2\ .
\ee
%
%
We may now consider a given $S$ characterised by
$p\mb$ and investigate a different trajectory specified by $q\mb$:
  \be
  \label{G3}
x\mo=x_0\mo+q\mo(\sigma-\sigma_0)\ .
\ee
For $q\mo q\mb=m^2$ it also
obeys the geodesic equation~\eqref{nov21-18-12}.
Evaluating the eikonal $S$~\eqref{G1}
along~\eqref{G3} yields
\be
\label{G4}
dT \equiv dS = 
p\mb q\mo d\sigma\ . 
\ee
This time measures the number of oscillations of the wave function~\eqref{G1} as counted on the trajectory for the reference frame $q\mb$.

Two different reference frames $q\mb$ and $q'\mb$ are related by a Lorentz transformation~\cite{CWPW}. This may be demonstrated by choosing a standing wave $p\mb=(m,0,0,0)$ and considering $q'\mb=p\mb$. One finds $p\mb q\mo = m\sqrt{\vec{q}^2 + m^2}$, while $p\mb q^{\prime\mu}=m^2$. With $dT\sim(p\mb q\mo)$, $dT'\sim(p\mb q^{\prime\mu})$ one obtains
\be
dT = \frac{\sqrt{\vec{q}^2 + m^2}}{m} dT^\prime =
  \frac{1}{\sqrt{1-v^2}} dT^\prime \; .
  \ee
For the second equation we note that the reference frame $q^{\prime\mu}$ moves relative to $q\mo$ with three-velocity $\vec{v}$, $v^2=\vec{v}^{\,2}$. The observer $q\mo$ \qq{sees} the ticks of the clock $q^{\prime\mu}$ with relativistic time dilation. \qq{Relativity} is obvious if we consider a different wave function for which $p\mo$ is replaced by $q\mo$. For this wave function the observer $q\mo$ measures proper time, while the observer $q^{\prime\mu}$ sees the same oscillations of the wave function with time dilation. We can generalise to two reference frames with arbitrary $q\mo$ and $q^{\prime\mu}$ in which the oscillations of the standing wave $p\mo$ are counted,
\bel{2.21A}
\frac{dT}{dT'}=\left(\frac{\vec{q}^2+m^2}{\vec{q}^{\prime2}+m^2}\right)^{1/2}\ .
\ee
This is the result of special relativity. Furthermore, our relations remain true if we apply the same Lorentz boost to $p\mo$, $q\mo$, and $q^{\prime\mu}$, since both $dT$ and $dT'$ involve only Lorentz-scalars. There is therefore no restriction in the choice of $p\mo$, $q\mo$, and $q^{\prime\mu}$.
We see that
our concept of defining physical time by counting the oscillations of the wave function generates in a natural way the concept of
relative time in special relativity. It is a simple way to
extract special relativity from basic quantities in (quantum) field theory.

We can generalise this concept to rather arbitrary
  trajectories exploring
  a given wave function $S$. This includes arbitrary $m(x)$, $g\mnb(x)$,
  arbitrary solutions $S(x)$, and arbitrary trajectories (with mild
  conditions).
  Inversely, consider a family of geodesic trajectories.
With $S(\sigma)$ computed for every geodesic trajectory
according to \eqref{sep17-22-2}, one can construct
a particular solution $S(x)$ of eq.~\eqref{nov21-18-3}
if every point $x^\mu$ of the manifold belongs to some
geodesic trajectory (no caustics,
  forbidden regions, etc.). This gives a solution to the
  Cauchy problem of eq.~\eqref{nov21-18-3}: the initial value
  $x^\mu (\sigma_0)$ is taken at the Cauchy hypersurface, so
  $S(\sigma_0)$ is determined by Cauchy data. As an example, one
may take a family of geodesic trajectories,  labeled by
${\bf x} (\sigma_0)$, with one and the same
initial time coordinate $x^0$ at $\sigma_0$.

\section{Homogeneous and isotropic Universe}

\subsection{World lines and geodesic physical time}

We now specify to a spatially flat FLRW  metric
\be
ds^2 = N^2(t) dt^2 - a^2(t) \delta_{ij} dx^i dx^j \; 
\label{oct30-17-5a}
\ee
and assume that the mass depends on time only, $m=m(t)$. The \qq{coordinate time} $t$ depends on the choice of $N(t)$. For $N=1$ it is \qq{cosmic time}, while for $N=a$ the coordinate time coincides with conformal time $\eta$. The coordinates and metric defined by~\eqref{oct30-17-5a} are the ones used for the field equation~\eqref{may12-20-1} and its solution. For the particular case of constant $Z$ and conformal time ($N=a, t=\eta, \mathcal H=\d_\eta\ln a$) the field equation~\eqref{may12-20-1} reads
\be\label{H1}
\gl\d_\eta^2+2\mathcal H\d_\eta-\delta^{ij}\d_i\d_j+a^2m^2\gr\phi=0\ ,
\ee
while the eikonal equation~\eqref{nov21-18-3} is given by
\be\label{H2}
\gl\d_\eta S\gr^2-\sum_i\gl\d_iS\gr^2-a^2m^2=0\ .
\ee
With
\be\label{H3}
\phi=\tilde\phi(\eta)e^{-ik_ix^i}\ ,\quad S=\tilde S(\eta)-k_ix^i\ ,\quad k^2=\delta^{ij}k_ik_j\ ,
\ee
one obtains
\be\label{H4}
\gl\d_\eta^2+2\mathcal H\d_\eta+k^2+a^2m^2\gr\tilde\phi=0\ ,
\ee
and
\be\label{H5}
\d_\eta\tilde S=\sqrt{k^2+a^2m^2}\ .
\ee
For the rapid oscillations of our interest both $\mathcal H\d_\eta$ and and $a^2m^2$ are small as compared to $k^2$. Both $\eta$ and $k^2$ as well as $a^2m^2$ are independent of the metric frame. On the other hand, the choice $Z=1$ is not frame invariant, neither is $\mathcal H$. In the limit $m\to0$, $\mathcal H\to0$ the solutions of the wave equation are plane waves,
\be\label{H6}
\phi=\phi_0\exp\left\{i\gl k\eta-\vec{k}\vec{x}\gr\right\}\ .
\ee
In this limit the eikonal approximation becomes exact with $A=\phi_0$.

For conformal time the world line associated to the particular solution~\eqref{H3},~\eqref{H5} of the wave equation is given by
\bel{C1}
u^0=\frac{1}{a^2}\deta S=\frac1{a^2}\sqrt{k^2+a^2m^2}\ ,\quad
u^i=-\frac{1}{a^2}\frac{\d S}{\d x^i}=\frac{1}{a^2}k_i\ .
\ee
With $x^0=\eta$, one infers
\bel{C2}
dT=\frac{a^2m^2}{\sqrt{a^2m^2+k^2}}d\eta\ .
\ee
For $am\to0$ generalised geodesic completeness requires
\bel{C3}
\int_{-\infty}^{\eta}d\eta a^2m^2=\infty\ ,
\ee
which is violated for $am\sim|\eta|^{-\alpha}$ if $\alpha>1/2$.

With $NdT=ad\eta$ the geodesic physical time interval \eqref{aug28-22-11} is given for the general metric~\eqref{oct30-17-5a} by
\be
d T = m^2 d\sigma  = \frac{aNm^2}{\sqrt{a^2 m^2 +k^2}} dt \; .
\label{aug28-22-30}
\ee
It is invariant under both
time reparametrization and Weyl
scaling. The Weyl-invariant generalization of
the criterion \eqref{aug17-22-1} of past geodesic completeness is
\be
\int_{t=-\infty}^t~dt~ \frac{aNm^2}{\sqrt{a^2 m^2 +k^2}} = \infty \; .
\label{aug28-22-20}
\ee
In the most interesting case when $a(t) \to 0$ as $t\to -\infty$,
eq.~\eqref{aug28-22-20} reduces to
\be
\int_{t=-\infty}^t~dt~ aNm^2  = \infty \; .
\label{aug28-22-20A}
\ee
Of course this criterion
agrees with  \eqref{aug17-22-1} for constant mass and $N=1$. A direct derivation of eqs.~\eqref{aug28-22-30}-~\eqref{aug28-22-20A} can be found in the appendix~\ref{appC}.

\subsection{Time-dependent mass in Minkowski space}

Let us consider Minkowski space and a $\phi$-particle with
time-dependent mass $m(t)$
such that $m(t) \to  0$ as $t \to -\infty$. This scenario may look somewhat special. It actually is not since many models for early cosmology, including inflation, can be formulated in a \qq{primordial flat frame} for the metric~\cite{Wetterich:2020fth}.
We recall
        eq.~\eqref{aug28-22-30} and see
        that the geodesic physical time 
        associated with such a particle is determined by
        \be
        T \propto \int_{-\infty}^t~dt~ \frac{m^2}{\sqrt{m^2 + k^2}} \; ,
\label{aug28-22-40}
        \ee
as the Minkowski time runs from $t=-\infty$ to finite $t$.
        If this integral is convergent, the setup is past
        time incomplete
        for the $\phi$-particle. 
        
To see what is going on, we first observe that the solution of the wave equation
\bel{W1}
\gl\d_t^2+k^2+m^2(t)\gr\phi=0
\ee
or eikonal equation
\bel{W2}
\d_t\tilde S=\sqrt{k^2+m^2(t)}
\ee
exists for any $k\neq0$ and $m(t)\to0$. It takes the simple limiting form~\eqref{H6} with $\eta=t$. In lowest order in $m/k$ one has
\bel{W3}
S=kt-\vec{k}\vec{x}+\frac1{2k}\int_{-\infty}^{t}dt'm^2(t')\ .
\ee

The wave front $S=0$, $\vec x=x_w\vec k/k$ obeys
\bel{W4}
x_w=t+\frac1{2k^2}\int_{-\infty}^{t}dt'm^2(t')\ ,\quad \dot{x}_w=1+\frac{m^2}{2k^2}\ .
\ee
On the other hand, the trajectory of a particle moving on the associated general geodesics is given by $\vec x=x_p\vec k/k=x_p\vec n$, cf. eq.~\eqref{may31-20-11},
\bel{W5}
\dot{x}_p=\frac k{\sqrt{k^2+m^2}}=1-\frac{m^2}{2k^2}\ .
\ee
As long as $m^2>0$ the particle is slower than the wave front, $\dot{x}_p<\dot{x}_w$. In the reference frame of this particle one can therefore count the oscillations of the wave front overtaking
it. These are precisely the ocsillations
 that determine the physical time in the particle rest frame.
For $m\to0$, however, the two velocities $\dot{x}_p$ and $\dot{x}_w$ coincide, such that in the particle rest frame the wave becomes frozen and the geodesic clock no longer ticks. This reflects, of course, the well known fact that for massless particles no rest frame can be defined. If the approach to the massless situation is too rapid, the number of ticks towards $t\to-\infty$ remains finite.

From the point of view of a hypothetical observer in the particle rest frame something dramatic must happen if geodesic time is incomplete. We may calculate
the acceleration $W$ in the inertial reference frame in which the
$\phi$-particle
is instantaneously at rest at time $t$ (instantaneously comoving
inertial frame)
and compare it
with $m (t)$.
We consider a one-dimensional motion with
$y=z=0$, $x = x(t)$.
The acceleration in the instantaneously comoving reference frame is
\be
W = \frac{\ddot{x}}{(1-\dot{x}^2)^{3/2}} \; .
\ee
In our case we have (see \eqref{may31-20-11})
\be
\dot{x} = \frac{k}{\sqrt{k^2 + m^2(t)}},
\ee
so 
\be
W = - k\frac{\dot{m}}{m^2}\ ,
\ee
and the ratio of interest is
\be
\frac{W}{m (t)} = - k \frac{\dot{m}}{m^3}\ .
\ee
If the integral \eqref{aug28-22-40} is convergent,
$m(t)$ decays backwards in time
faster than $|t|^{-1/2}$ and
\be
\left|\frac{W}{m (t)}\right| \to \infty \;\;\;\; \mbox{as} \;\;
t\to -\infty \; .
\ee
Thus, the $\phi$-particle experiences infinite acceleration
in its own reference frame as $t\to -\infty$; in particular,
one would not be able to construct any
clock using $\phi$-particles.

Any spatially flat cosmological metric can be Weyl scaled 
to Minkowski space (flat metric frame),
as emphasized recently, e.g., in
Ref.~\cite{Wetterich:2020fth}. Our simple exercise shows how the geodesic incompleteness of a geometry
in the metric frame where $m=\mbox{const}$ translates to the flat metric frame. 

\section{Cosmic reference frame and massless particles}\label{sec:CMP}

\subsection{Cosmic reference frame}

For a homogeneous isotropic metric we can define the particular trajectory
\bel{CR1}
x^0=c\sigma\ ,\quad x^i=\text{const.}
\ee
This defines the \qq{cosmic reference frame}. As a convenient choice we identify $x^0$ with conformal time $\eta$, which is a Weyl-invariant quantity. For the solution of the wave function~\eqref{H3}-~\eqref{H5} the counting of oscillations becomes rather simple, with physical time given by
\bel{CR2}
\Delta T=\Delta S=\int_{\eta_{in}}^{\eta}d\eta\partial_\eta\tilde S(\eta)=\int_{\eta_{in}}^{\eta}d\eta\sqrt{k^2+a^2m^2}\ .
\ee

For solutions with $k\neq0$ this expression always diverges for $\eta_{in}\to-\infty$. The clocks defined by the cosmic reference frame tick then an infinite number of times. The corresponding clock system is \qq{time complete}. The limits $a\to0$ or $m\to0$ pose no particular problem. In particular, for $am\sim|\eta|^{-\alpha}$, $\alpha>1/2$, the \qq{cosmic physical time} defined by the cosmic reference frame is infinite, while it is finite for geodesic physical time, c.f. eq.~\eqref{C3}. We conclude that in this case cosmic physical time and geodesic physical time constitute inequivalent clock systems.

We can define cosmic physical time for solutions of the wave equation for massless particles as photons or gravitons. For $m=0$ eq.~\eqref{CR2} yields
\bel{CR3}
T=k\eta\ .
\ee
The clocks for solutions with different $\vec k$ all belong to the same equivalence class. For wave functions of massless particles we can identify cosmic physical time with conformal time.

Geodesic physical time has the important advantage that it can be defined for arbitrary metrics and arbitrary $m(x)$. A similar general definition of cosmic physical time is not obvious. For rather arbitrary geometries photons and gravitons are massless and their wave functions are oscillatory. One would expect that there are many trajectories or reference frames along which the number of oscillations towards the past extends to infinity. The problem concerns the selection of a natural cosmic reference frame. Conformally flat geometries can be mapped by Weyl transformations to a flat metric. Since conformal time is invariant under Weyl transformations we can use it as cosmic physical time for this class of metrics. For more general inhomogeneous metrics one may think of a cosmic reference frame associated to a suitable average of the metric. Indeed, conformal time remains a useful concept for practical cosmology where geometry is described by small deviations from a homogeneous isotropic background. In an intuitive sense, the detailed geometry should play no role for the oscillations with wave lengths much shorter than all characteristic length scales of the geometry. A precise formulation of cosmic physical time along this concept needs to be worked out, however.

\subsection{World with massless particles}

One may wonder if physical time and time completeness can be defined in a world with only massless particles. For $m=0$ geodesic physical time does not exist, while cosmic physical time remains a possibility. The standard way of deriving the criterion \eqref{aug17-22-1} is to
make use of the affine parameter of massless geodesics~\cite{Borde:2001nh}.
However, the issue of geodesic (in)completeness in a world where all particles
have strictly zero masses (``photons'')
is controversial. On the one hand, photons behave in FLRW space-time
\eqref{aug28-22-60} exactly like in Minkowski space with coordinates
$\eta, {\bf x}$. So, if the cosmological evolution starts at conformal time
$\eta = -\infty$, one is tempted to suggest that any such cosmology
is complete. This viewpoint has been advocated, e.g., in
Ref.~\cite{Wetterich:2020fth}. According to it, the photon
physical time
is determined by
the number of oscillations of the photon wave function
in the homogeneous and isotropic reference frame. This number is indeed
proportional to conformal time $\eta$ and corresponds to cosmic physical time.

For a massless scalar field a field redefinition
\be
\phi (\eta, {\bf x})  = \lambda (\eta) \hat{\phi} (\eta, {\bf x}) 
\ee
transforms the action in
a cosmological background (in
conformal time)
\be
\mathcal{S} = \frac{1}{2} \int~d\eta ~d^3x a^2(\eta) \eta^{\mu \nu}
\d_\mu \phi \d_\nu \phi
\ee
(where $\eta^{\mu \nu}$ is Minkowski tensor) into
\be
\mathcal{S} = \frac{1}{2} \int~d\eta ~d^3x \lambda^2 (\eta)
a^2(\eta) \eta^{\mu \nu}
\d_\mu \hat{\phi} \d_\nu \hat{\phi} + \dots
\ee
where omitted terms are irrelevant in the lowest WKB order.
Since $a$ and $\lambda a$ should be equivalent this reiterates that the  behavior of the scale factor
as function of time should be irrelevant for strictly massless
particles.

On the other hand, a general cosmic reference frame is not (yet) available). It remains to be understood what (if anything) replaces the
notion of geodesic (in)completeness in a world where all particles have
strictly zero mass.

\section{Conclusions}\label{sec:C}

We propose to generalise the purely geometric concept of geodesic completeness to \qq{time completeness}. Time completeness is based on physical time as defined by counting the number of oscillations of a wave function or some other (quasi-) periodic process. Physical time and the associated notion of completeness is independent of the choice of the metric frame. For particles with nonzero, but possibly varying mass $m$ a frame invariant \qq{geodesic physical time} interval $dT$ can be defined as the product of proper time with mass $dT=md\tau$. This dimensionless quantity counts the number of oscillations of the wave function of a massive particle, as measured on a particular world line or generalised geodesic derived from this wave function. In the limit of constant $m$ time completeness is equivalent to the geometric notion of geodesic completeness if geodesic physical time is used.

Physical time is not defined uniquely. Different trajectories on which oscillations of a given wave function are counted are associated to clocks in different reference frames. In case of flat geometry and constant $m$ these reference frames are related by Lorentz transformations. Our setting relates physical time for rather arbitrary reference frames. Different wave functions define different clocks for physical time as well.

Often clocks are equivalent in the sense that each tick of one clock is mapped to a tick of the other clock. This does not hold for all clocks, however, as we demonstrate for a \qq{cosmic reference frame} that defines a complete physical time from wave functions of particles with arbitrary mass, including the massless case. For the same wave function geodesic physical time may not be complete. We conclude that time completeness is robust in the sense that it does not depend on the metric frame and is directly related to possible physical processes. It is not universal, however, and one may debate which choice of physical time is most useful for an understanding if a given cosmological model needs an extension or not.

\vspace{1cm}

Acknowledgement: Our collaboration has been interrupted at a very final stage by the tragic death of Valery R. He could not do the \qq{final wording} that was foreseen before publication. V.R. would like to thank P. Creminelli for fruitful discussions.

\begin{appendices}
  \section{Derivation of the geodesic equation with varying mass}
  \label{appA}

  Consider a partial differential equation for an unknown function $S$
  (action, eikonal):
\be
F(x^\mu , p_\nu) = 0
\; , \;\;\;\;\; \mbox{where} \;\;\; p_\mu = \d_\mu S \; .
\ee
Let $S (x)$ be a solution. Then
\be
\frac{\d F}{\d x^\mu} + \frac{\d F}{\d p_\lambda} \cdot
\frac{\d p_\lambda}{\d x^\mu} = 0 \; .
\label{nov21-18-1}
\ee
Now,
\be
\frac{\d p_\lambda}{\d x^\mu} = \frac{\d^2 S}{\d x^\lambda \d x^\mu}
= \frac{\d p_\mu}{\d x^\lambda} \; .
\ee
Therefore, eq. \eqref{nov21-18-1} reads
\be
\frac{\d F}{\d x^\mu} + \frac{\d F}{\d p_\lambda} \cdot
\frac{\d p_\mu}{\d x^\lambda} = 0 \; .
\label{nov21-18-2}
\ee
Consider a line (characteristic)
parametrized by $\sigma$, such that
\be
\frac{d x^\lambda}{d \sigma} = \frac{\d F}{\d p_\lambda}\; .
\ee
Along this line
\be
\frac{d p_\mu}{d\sigma} = \frac{\d p_\mu}{\d x^\lambda}\cdot
\frac{d x^\lambda}{d \sigma} =  \frac{\d F}{\d p_\lambda} \cdot
\frac{\d p_\mu}{\d x^\lambda} =
- \frac{\d F}{\d x^\mu} \; .
\ee
Thus, we come to the system of Hamilton
equations for the characteristic:
\begin{align}
  \frac{d x^\lambda}{d \sigma} &= \frac{\d F}{\d p_\lambda} \; ,
  \\
  \frac{d p_\mu}{d \sigma} & = - \frac{\d F}{\d x^\mu} \; .
\end{align}
The increment of
the action (eikonal) along this line is found from
\be
\frac{dS}{d \sigma} = p_\mu \frac{d x^\mu}{d\sigma}
= p_\mu \frac{\d F}{\d p_\mu}
\ee
In the specific case \eqref{nov21-18-3} we have
\be
F = \frac{1}{2} \left[ g^{\mu \nu}(x) p_\mu p_\nu - m^2(x) \right]
\ee
Hence, the Hamilton equations are
\begin{align}
  \frac{d x^\lambda}{d \sigma} &\equiv u^\lambda = g^{\lambda \mu}p_\mu \; ,
  \\
  \frac{d p_\mu}{d \sigma} &= - \frac{1}{2} \frac{\d g^{\nu \lambda}}{ \d x^\mu}
  p_\nu p_\lambda + \frac{1}{2} \frac{\d m^2}{\d x^\mu} \; .
\end{align}
These equations combine into the geodesic equation for $u^\mu$:
\be
\frac{d}{d \sigma} \left(g_{\mu \nu}(x) u^\nu \right)
+ \frac{1}{2} \frac{\d g^{\nu \lambda}}{\d x^\mu} g_{\nu \rho}
g_{\lambda \kappa} u^\rho u^\kappa - \frac{1}{2} \frac{\d m^2}{\d x^\mu}
=0 \; .
\label{nov21-18-11}
\ee
A straightforward manipulation gives eq.~\eqref{nov21-18-12}.
\label{nov21-18-10}

\section{Weyl invariant form of the geodesic\\equation}
\label{appB}

To cast the geodesic equation \eqref{nov21-18-12} in a form invariant under
the Weyl scaling \eqref{nov21-18-20}, \eqref{aug28-22-1}, we parametrize the
geodesic by the physical time $T$ instead of the parameter $\sigma$.
We introduce
\be
w^\mu = \frac{d x^\mu}{d T} = \frac{u^\mu}{m^2 (x)}
\ee
and obtain from eq.~\eqref{nov21-18-12}
\be
\frac{d w^\mu}{d T} +
\frac{2}{m} \frac{\d m}{\d x^\nu} w^\mu w^\nu +
\Gamma^\mu_{\nu \lambda} w^\nu w^\lambda
- \frac{1}{m^3} g^{\mu \nu}  \frac{\d m}{\d x^\nu} = 0 \; .
\label{nov21-18-13}
\ee
We recall that the parameter $T$ is invariant under the Weyl scaling and
find that
this equation is  Weyl invariant. Indeed, we use
\eqref{nov21-18-20}, \eqref{aug28-22-1},
recall that
\be
\Gamma^\mu_{\nu \lambda} =  \hat{\Gamma}^\mu_{\nu \lambda}
+\delta^\mu_\nu \d_\lambda \ln \Omega
+\delta^\mu_\lambda \d_\nu \ln \Omega
-\hat{g}_{\nu \lambda} \hat{g}^{\mu \rho} \d_\rho \ln \Omega
\ee
and that $g_{\mu \nu} w^\mu w^\nu = m^{-2}$, and find that
eq. \eqref{nov21-18-13} has the same form in variables with hats.

\section{Physical time and generalised completeness for general homogeneous isotropic metric}\label{appC}

In this appendix we display a direct derivation of the general expressions for $dT$~\eqref{aug28-22-30} and geodesic completeness~\eqref{aug28-22-20},~\eqref{aug28-22-20A}. We first compute the world lines for general $N(t)$ and $m(t)$.
The $0$-component of the geodesic equation
\eqref{nov21-18-12}
reads
\be
u^0\frac{du^0}{dt} + \left( \frac{\dot{N}}{N}
+ \frac{a \dot{a}}{N^2} \dot{x}^i \dot{x}^i \right) (u^0)^2
- \frac{1}{2} \frac{\d m^2}{\d t} \frac{1}{N^2} = 0 \; ,
\ee
where dot denotes $d/dt$;
we use the fact that $d/ d\sigma = u^0 \cdot d/dt$.
Using~\eqref{XA}, $g_{\mu \nu}u^\mu u^\nu = m^2(x)$, yields
\be
\dot{x}^i \dot{x}^i = \frac{N^2}{a^2} - \frac{m^2}{a^2 (u^0)^2} \; .
\label{aug28-22-10}
\ee
In this way we arrive at an equation for $(u^0)^2$:
\be
\frac{1}{2} \frac{d[(u^0)^2]}{dt}
+ \left( \frac{\dot{N}}{N} + \frac{\dot{a}}{a} \right)(u^0)^2
    - \frac{1}{N^2} \left( \frac{m^2 \dot{a}}{a}
    + \frac{1}{2} \frac{\d m^2}{\d t}
    \right) = 0\ .
    \ee
    The solution to this equation is
    \be
u^0 = \frac{\sqrt{m^2 a^2 + C}}{aN}\ ,
\label{may12-20-2}
\ee
 where $C$ is an arbitrary constant.
Also, we find from \eqref{aug28-22-10}
\be
\dot{x}^i =  \frac{N}{a} \frac{1}{\sqrt{m^2 a^2 + C}}
  \sqrt{C}n^i \; , \;\;\;\; n^i n^i =1 \; ,
\label{may31-20-11}
  \ee
  which relates the parameter $\sqrt{C}$ to particle
  velocity. The physical time interval~\eqref{aug28-22-30} follows from the definition~\eqref{aug28-22-11}, $dT=m^2d\sigma=\gl m^2/u^0\gr dt$, where $k^2$ is identified with the integration constant $C$.

\end{appendices}

\end{document}